\documentclass[letterpaper,prd,twocolumn,unsortedaddress]{revtex4}
\usepackage{graphicx}
\usepackage{epsfig}
\usepackage{hyperref}

\begin{document}

\title{Chiral Lattice Gauge Theories from 
Warped   Domain Walls
and Ginsparg-Wilson Fermions}
\author{Tanmoy Bhattacharya}
\email{tanmoy@lanl.gov}
\author{Matthew R Martin}
\email{mrmartin@lanl.gov}
\affiliation{T-8, MS B285\\ Los Alamos National Laboratory\\ Los Alamos, NM 87545 USA}
\author{Erich Poppitz}
\email{poppitz@physics.utoronto.ca}
\affiliation{Department of Physics\\ University of Toronto\\ Toronto, ON M5S 1A7 Canada}

\begin{abstract}
We propose a construction of a 2-dimensional lattice chiral gauge theory.  The construction may be viewed as a particular limit of an infinite warped 3-dimensional theory.  We also
 present a ``single-site'' construction  using Ginsparg-Wilson fermions which may avoid, in both 2 and 4 dimensions, the problems of waveguide-Yukawa models.
\end{abstract}

\maketitle

\section{Introduction and summary}

Understanding the strong-coupling behavior of chiral gauge theories is an outstanding problem of great interest, 
both on its own and for its possible relevance to phenomenology: the Standard Model of elementary particle physics is a chiral gauge theory and additional strong  chiral gauge dynamics at the (multi-) TeV scale may be responsible for breaking the electroweak symmetry and fermion mass generation. 

The only  clues of the strong coupling behavior of non-supersymmetric chiral gauge theories come from 't Hooft anomaly matching and most attractive channel arguments. Large-N expansions, including the recently considered gravity duals in the AdS/CFT(QCD) framework, do not apply to chiral gauge theories. Thus, the space-time lattice regularization remains, to this day, the only way to advance our limited knowledge of  chiral gauge dynamics. 

The lattice, however, fails to reproduce the physics of a chiral gauge theory due to the presence of extra, unwanted fermion ``doubler" modes. The difficulty of this problem is encoded in a no-go theorem \cite{Nielsen:1981hk}.  Recent reviews of the different approaches to lattice chiral gauge theories are  \cite{Golterman:2000hr} and  \cite{Luscher:2000hn}; see also ref.~\cite{Golterman:2004qv} for a new   approach.

It has long been known that the extra fermions \emph{can} be removed from the spectrum by sacrificing the gauge symmetry, at least perturbatively \cite{BrokenChiralTheory, Borrelli:1989kw}.
Recently, a proposal was made to do that in a new way~\cite{Bhattacharya:2005xa, Bhattacharya:2005mf}. 
 The fermion masses must be chosen in a non-trivial way to break the appropriate global symmetries in order to reproduce the anomalies of the target theory, while still maintaining the appropriate light fermion modes.  The gauge symmetry can then be restored through a limiting process inspired by a 5-dimensional model in Anti-de Sitter space.  Unfortunately for this approach, called ``warped domain wall fermions,'' the associated Goldstone mode is strongly coupled leaving the model's status somewhat uncertain. 

The warped domain wall fermion model bears some similarity to the ``wave-guide model''~
\cite{Kaplan:1992bt,Kaplan:1992sg}, which is known not to give a chiral theory \cite{Golterman:1993th,Golterman:1994at}.  There are, however, significant differences between the models---in particular, in the warped case the source of the gauge boson and fermion mass are decoupled and so further investigation of this model is necessary.  

In this paper, we first consider a  construction analogous to that of~\cite{Bhattacharya:2005xa} in 3-dimensional Anti-de Sitter space in an attempt to construct a 2-dimensional chiral gauge theory.  We describe a limit which results in a 2-dimensional chiral gauge theory without a strongly coupled Goldstone mode.  

We then propose a related, simplified ``one-site" model which consists of only a 2-dimensional  lattice theory where massless fermions are introduced using the Ginsparg-Wilson mechanism~\cite{Ginsparg:1981bj,Luscher:1998pq} for imposing a modified chiral symmetry. The symmetries and anomalous Ward identities in this model are as expected in the target theory. Furthermore, a preliminary  strong coupling analysis (a more detailed study of this model is in progress), which is also expected to hold in the 4 dimensional version of the construction, indicates that no new unwanted light fermions appear and the fermion spectrum of the unbroken gauge theory remains chiral in an appropriately taken limit.

In addition to providing some insight on the workings of the 4-dimensional warped domain wall fermion construction, the models presented here are interesting for their own sake as they are the simplest examples of chiral gauge theories. It is also clear that 2-dimensional models are   the most amenable to  numerical tests and to this end alone it is  desirable to have an appropriate formulation; in the process,  we will see that the warped 2-dimensional case presents a number of subtleties compared to the  4-dimensional warped domain wall fermions. 

This paper is organized as follows. We begin in  Section II,  where we describe previous work on lattice chiral gauge theories within the ``wave-guide model." We review the earlier arguments showing that the wave guide model gives rise to a vectorlike spectrum of massless fermions, both at small and large Yukawa coupling.

In Section III, we  explain how the proposal of \cite{Bhattacharya:2005xa} addresses the difficulty with obtaining a chiral spectrum using a warped $AdS$ background. Motivated by the strong  fermion/goldstone mode coupling found in the 4-dimensional implementation of the proposal, we construct and  study the much simpler 2-dimensional version in detail. Within perturbation theory and in the deconstructed description, we show that there are no bulk goldstone/fermion strong interactions in this model and that the desired spectrum and separation of scales can be achieved while at weak coupling (two appendices describe various important technical details). 
Our results of Section III indicate that the "warped domain wall" framework for lattice chiral gauge theories is still of interest and deserves further study, including a full lattice implementation. 

In  Section IV, we  present another  proposal: the "single-site model." It is related to the "warped domain wall" in that it   is also motivated  by considering the waveguide model and its failure to give a chiral fermion spectrum, this time in the strong Yukawa coupling regime. Our  main observation here is that using Ginsparg-Wilson fermions helps avoid the left/right mixing that leads to a vectorlike fermion spectrum. To also obtain a massless gauge boson, we have to make use of the strong-Yukawa symmetric phase of the Yukawa-Higgs theory. We give a plausibility argument as to why we believe this phase can be  realized in our construction without fine tuning. Further analytical and numerical work on the "one-site" model supporting our proposal will appear in \cite{GP2006}.
 
 In Section V, we conclude with a summary of the  proposals, a list of outstanding issues, and an outlook for future work.

\section{Domain Wall Fermions in Two Dimensions}\label{sec:2dDomainWall}
We review here some features of fermions in two dimensions and discuss some relevant previous work on chiral gauge theories.  By understanding the short-comings of previous attempts we will see how our construction differs in important ways.

A 2-dimensional Dirac fermion has two complex components:
\begin{equation}
\Psi \equiv \left( \begin{array}{c} \psi_- \\ \psi_+ \end{array}\right).
\end{equation}
We will call the $\psi_-$ field left handed and the $\psi_+$ field right handed.  We work with light-cone coordinates:
\begin{equation}
x^{\pm} \equiv t \pm x \quad \to \quad 2 \partial_{\pm} = \partial_t \pm \partial_x,
\end{equation}
so that the Lagrangian for a charged, massive, Dirac fermion is:
\begin{eqnarray}
{\cal L} &=& 2i \; \bar\psi_- \left( \partial_+ - i A_+ \right) \psi_- + 
2i \; \bar\psi_+ \left( \partial_- - i A_- \right) \psi_+  \nonumber \\
&& \quad + m_D \bar\psi_+ \psi_- + m_D^* \bar\psi_- \psi_+ .
\end{eqnarray}
The bar on these one-component, complex fields indicates complex conjugation, while the subscript $D$ on the mass indicates that it is a Dirac type mass term: it does not break a gauge symmetry.  We will later introduce masses of the Majorana type which would break gauge symmetry:  $m_M \psi_+ \psi_- + \textrm{h.c.}$.  By Lorentz invariance there are no other types of mass terms: any mass must couple left and right handed fermions.  This restriction of mass terms will be important later.

The domain wall fermions~\cite{Kaplan:1992bt} arise from consideration of a theory with a third dimension labeled with the coordinate $z$.  With the appropriate mass terms there will be a light left-handed mode localized at one end of the extra dimension and a light right-handed mode localized at the other end.  We will consider the theory with the $z$ direction on a lattice, keeping the other two directions as a continuum.  Upon discretization the kinetic term in the $z$ direction contributes to mass and mixing terms for the fermions on adjacent sites:
\begin{equation}
\bar\psi \partial_z \psi \, \to \, \frac{\bar\psi_i \left(\psi_i - \psi_{i-1}\right)}{\delta z}.
\end{equation}
This is sometimes referred to as the deconstructed model~\cite{Arkani-Hamed:2001ca}.  Alternatively we may think of this as a lattice in all three directions where the lattice spacing in the two $x^\mu$ directions is much smaller than the lattice spacing in the $z$ direction.  Indeed, we will never take the $z$ lattice spacing to zero.

\begin{figure*}
\includegraphics{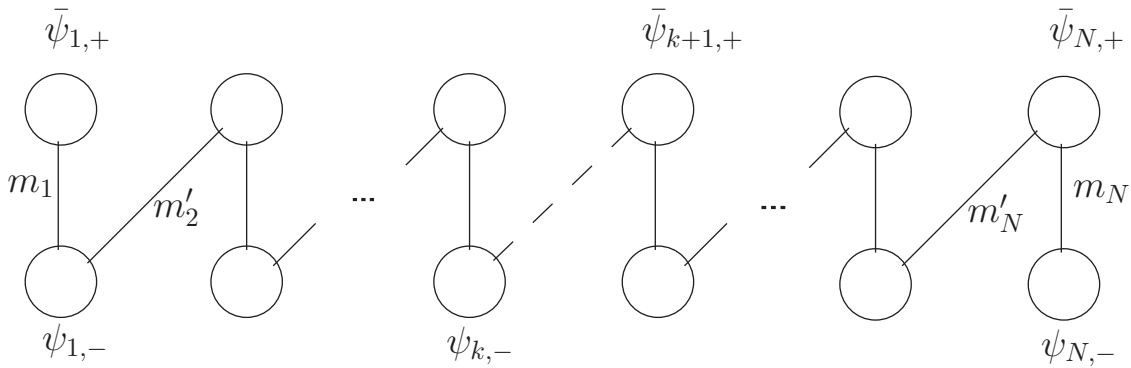}
\caption{The wave guide approach to a chiral gauge theory.  Circles represent Weyl fermions.  Solid lines represent mass terms and a charged scalar couples the charged fermion, $\bar\psi_{k+1,+}$, to the neutral fermion, $\psi_{k,-}$.\label{fig:WaveGuide}}
\end{figure*}

If we place two domain wall lattices back-to-back with one being charged and one being neutral we have the ``wave guide'' approach.  The Lagrangian is given by:
\begin{eqnarray}
&& \sum_{i=1}^k \Big[ 2 i\; \bar\psi_{i-} \partial_+ \psi_{i-} + 
2i \; \bar\psi_{i+} \partial_-  \psi_{i+}
+ \left( m_i \bar\psi_{i-}\psi_{i+} +\textrm{h.c.} \right) \Big] \nonumber \\
&& + \sum_{i=2}^k \left( m_i' \bar\psi_{i+} \psi_{i-1,-} + \textrm{h.c.} \right) \\
&& + \sum_{i=k+1}^N \Big[2i \;  \bar\psi_{i-} \left( \partial_+ - i A_+ \right) \psi_{i-} + 
2 i \; \bar\psi_{i+} \left( \partial_- - i A_- \right) \psi_{i+}  \nonumber \\
&& + \left( m_i \bar\psi_{i-}\psi_{i+} +\textrm{h.c.} \right) \Big] 
+ \sum_{i=k+2}^N \left(m_i' \bar\psi_{i+} \psi_{i-1,-} + \textrm{h.c.} \right) \nonumber
\end{eqnarray}
and schematically, this Lagrangian is represented in figure~\ref{fig:WaveGuide}.
The masses $m_i$ and $m_i'$ depend upon the details of the discretization and are not important at this point.  Note that the gauge mode is independent of the sites: we are treating it as a 2-dimensional degree of freedom.  In order to couple the charged and uncharged fermions in a gauge invariant manner we introduce a charged scalar, $\phi$, and the coupling:
\begin{equation}
\label{yukawa}
y \bar\psi_{k+1,+} \phi \psi_{k,-} + \textrm{h.c.}.
\end{equation}
The phases of the analogous
model in both two and four dimensions were analyzed for both weak and strong Yukawa couplings $y$ in~ \cite{Golterman:1993th},\cite{Golterman:1994at}, with the conclusion that the theory is non-chiral in every case. The
simplest possibility is for $y =0$. In this case one can easily
see that the model falls apart into two
disconnected theories. One is the fully gauged wave guide part and the
other is the ungauged part of the domain wall.  Each of these two
parts themselves form a domain wall model, and
each of these will either have zero modes localized at both ends or at
neither end.  Thus the boundary of the wave guide will act as a domain
wall boundary itself.  Nothing qualitatively different happens
for small non-zero Yukawa, as long as the field
$\phi$ does not acquire a VEV.

However, if the scalar does obtain a VEV, then the light 
fermion mode localized at the wave guide
boundary could be eliminated using the opposite chirality fermion
localized on the other side of the wave guide boundary via the mass term
$y \psi_{k+1,+} \langle \phi \rangle \psi_{k,-}$. The problem with this 
approach is rooted in the fact that the gauge field does not fluctuate 
in the extra dimension.
The fermion mass obtained this way will be of the order $m_f \sim
y \langle \phi \rangle$. However, in this Higgs' mechanism, the
gauge boson will also pick up a mass of order $m_0 \sim g \langle \phi
\rangle$.  To get to an unbroken chiral theory one would like $m_0 \ll
m_f$, however their mass ratio is given by $g/y$. 
Since in four dimensions the Yukawa is an IR free coupling, at low energies
its value will be determined by $g$, and it seems that no hierarchy
between the masses is possible in the weak-coupling region. In the next section, we will explain how
 the ``warped domain wall" proposal avoids this problem by separating the scales of gauge boson and fermion masses.

In the opposite limit of   strong Yukawa coupling, the  phase structure of the $g=0$ lattice Higgs-Yukawa model with a fixed-length Higgs field was also analyzed in \cite{Golterman:1994at} via a strong coupling expansion in $y$ and it was again found that the spectrum of the model was vectorlike. This is most easily seen at leading order in $1/y$ by first rescaling the fermion fields at the boundary of the waveguide, see eqn.~(\ref{yukawa}), by $1/\sqrt{y}$, thus making their kinetic terms vanish  at $y\rightarrow \infty$. In our deconstructed picture of fig.~1 this results in removing the two circles adjacent to the dashed line; thus, after the rescaling,  the remaining charged $\Psi_{k+1, -}$ loses its Wilson term in the 2d noncompact lattice directions not shown---recall that the Wilson term couples $\Psi_{k+1, -}$ to the now absent $\Psi_{k+1, +}$. Naturally, the loss of the Wilson term  results  in the appearance of a plethora of charged and neutral massless states near the waveguide boundary,   localized at the new boundaries of the split waveguide, leading once more to a vectorlike spectrum \cite{Golterman:1994at}.
Thus,  it was concluded that  also in the strong Yukawa limit it is not possible to get a chiral gauge
theory from domain wall fermions. 
In section IV, we will explain how using the Ginsparg-Wilson mechanism of imposing a modified chiral symmetry on the lattice 
avoids the mixing of  light and mirror modes in the Yukawa coupling that led to the appearance of doublers \cite{Golterman:1994at}.

In~\cite{Bhattacharya:2005xa} it was argued that the situation is different when one allows the gauge field to fluctuate in the extra dimension and when one is considering a non-trivial background metric along the extra dimension.  In this case the scaling of the gauge boson mass could be different from that of the fermion mass in the presence of a symmetry breaking VEV on one of the domain wall boundaries. This led to a possibility of recovering a chiral gauge theory in the limit when the warping (the background curvature of the extra dimension) is increased to infinity.  We will repeat much of that argument below in the context of a 2-dimensional domain wall theory.

\section{A Warped 3-dimensional Theory}\label{sec:WarpedTheory}

\subsection{Gauge Fields}\label{sec:GaugeFields}
The key feature of this construction, as in the 4-dimensional case~\cite{Bhattacharya:2005xa}, is the separation of scales which is made possible by an appropriate introduction of curvature.  It is known that in a theory with a compact extra dimension and gauge symmetry breaking at one boundary, the mass of the gauge boson is not set by the VEV of the Higgs field alone~\cite{Huber:2000fh}.  The lightest gauge mode bends in the extra dimension and the associated gradient terms also contribute to the mass.  In the limit that the Higgs VEV goes to infinity, the gauge field is repelled from the location of symmetry breaking and the mass of the gauge boson is independent of the VEV.  In flat space this mass is parametrically the same as the masses of the Kaluza-Klein modes, but in a warped background there is the possibility of a separation of the lightest mode from the KK modes.  We chose our warping and boundary conditions so as to achieve that separation.

In order to study the strong coupling of a 2-dimensional chiral gauge theory, we require a hierarchy not just between the KK modes and the mass of the lightest gauge boson, but we must have the strong coupling scale of the theory lie between these two scales:
\begin{equation}
m_{KK} \gg \Lambda_{\chi GT} \gg m_{A_0}. \label{eq:MassScaleHierarchy}
\end{equation}
The scale $m_{KK}$ is the scale at which all of the 3-dimensional physics enters.  Much below this scale we are left with a 2-dimensional theory, as demonstrated by some simple checks of the mass spectrum and in appendix~\ref{sec:BetaFunc}.  The mass of the lightest gauge mode, $m_{A_0}$, sets the scale of the gauge symmetry breaking, and much above this scale the theory has an unbroken gauge symmetry.  The low energy 2-dimensional gauge theory has a gauge coupling, $g_2$, with units of mass, and therefore the gauge coupling itself sets the scale of strong coupling for this gauge theory.  At tree level in the 2-dimensional theory, we have $\Lambda_{\chi GT} = g_2$.

We first describe our construction in terms of a continuum theory living in a slice of 3-dimensional Anti-de Sitter space, $AdS_3$.  The metric is given by:
\begin{equation}
\label{interval}
ds^2 = \left(\frac{R}{z}\right)^2\left(\eta_{\mu\nu}dx^\mu dx^\nu - dz^2\right),
\end{equation}
where $x^\mu$ are the two flat directions and $z$ is the extra, warped direction.  The space is bounded in the $z$ direction.  One end, $z=R$, is called the UV brane.  The other end, $z=R'$, is called the IR brane, where $R' \gg R$.  We will break the gauge symmetry with a Higgs mechanism on the UV brane.  Note that this is different from~\cite{Bhattacharya:2005xa} and many phenomenological models~\cite{HiggslessPheno}, where the gauge symmetry is broken at the IR end.  The reason for the difference is that the scaling of the mass of the lightest gauge boson depends crucially on the number of dimensions as can be seen below.

The gauge field action is:
\begin{eqnarray}
&&\int d^3x \sqrt{g} \Big[- \frac{1}{4 g_3^2} F_{MN} F^{MN} \label{eq:AdS3Action} \\
&& \qquad + \delta(z-R)\left(\frac{1}{2}D_\mu \phi^* D^\mu \phi - V(\phi) \right)\Big], \nonumber
\end{eqnarray}
where $\phi$ is a UV-brane localized Higgs field, which we will take below to have a fixed VEV. Then, 
the bulk equation of motion for the KK modes of the transverse components of the gauge field is: 
\begin{equation}
\label{KKeqn}
\frac{R}{z} \partial_z \left( \frac{z}{R} \partial_z f_n(z)\right) = -m_n^2 f_n(z),
\end{equation}
where $m_n$ is the 2-dimensional mass squared of the $n$-th KK mode.
The solutions for the KK modes are Bessel functions:
\begin{equation}
f_n(z) = A_n J_0(m_n z) + B_n Y_0(m_n z). \label{eq:besselWaveFunctions}
\end{equation}
The boundary conditions also come from requiring that the boundary terms in the variation of the action vanish;  without boundary terms in the action, the allowed boundary conditions are Dirichlet or Neumann. By choosing the Higgs VEV large enough, we have effectively Dirichlet boundary conditions at the UV end  (see~\cite{HiggslessPheno}). We   choose Neumann boundary conditions  at the IR end because we want the gauge group to be unbroken there.  This leads to a mass spectrum well approximated by: 
\begin{equation}
\label{gaugeKK}
m_{A_n} \sim {n\pi\over R'}  \quad \to \quad m_{KK}\equiv \frac{\pi}{R'} ~,\label{eq:KKGaugeBoson}
\end{equation}
except for the lightest mode which has a mass:
\begin{equation}
\label{gaugelight}
m_{A_0}^2 = \frac{2}{R'^2} \frac{1}{\ln(R'/R)}\left(1+ {\cal O} \left(\frac{1}{\ln(R'/R)}\right)\right).
\label{eq:LightGaugeBoson}
\end{equation}
We can see right away that the physics of the 3-dimensional theory, set by the Kaluza-Klein scale $m_{KK}$ can be separated from the physics of the gauge symmetry breaking for large $\ln(R'/R)$.

The powers of the ratios of $R/z$ in the KK equation (\ref{KKeqn}) depend on the dimensionality of the space: had this been a slice of $AdS_5$, then both $R/z$ factors would have been inverted.  By breaking the gauge symmetry on the UV brane we find the same scaling for the mass of the gauge boson as in 5-dimensions with IR brane breaking \footnote{In fact, it is easy to see that the mass spectrum of the entire spin-1 KK tower in $AdS_3$ is identical to that in  $AdS_5$ with reversed boundary conditions; we thank T. Gherghetta for discussions.}. 

For the deconstruction description we choose a small dimensionless lattice spacing parameter, $a$, and let the physical lattice spacing scale across the space: $\delta z_i \sim  z_i$, ensuring that the interval (\ref{interval}) $\delta z_i/z_i$ between two neigboring lattice points is $i$-independent; eqn.~(\ref{warpfactor}) below implicitly defines the exact expression for  $\delta z_i$.  
It is helpful to define a local warp factor which encodes how much the metric warps from one lattice site to the next:
\begin{equation}
\label{warpfactor}
w = \frac{1}{1+a}  \, \to \, z_i = w^{1-i}R.
\end{equation}
If there are $N$ slices, then $R' \equiv z_N =  w^{1-N}R$.

We may consider the deconstructed Lagrangian coming from equation~(\ref{eq:AdS3Action}) in the $A_z=0$ gauge.  Alternatively this may be viewed as the Lagrangian for the $N$, coupled, 2-dimensional gauge theories in unitary gauge.  The UV boundary Higgs is left in:
\begin{eqnarray}
&& - \frac{1}{4} \sum_{i=1}^N \frac{a z_i}{R g_3^2} \left[ F^{i}_{\mu\nu}\right]^2
+ \frac{1}{2}D^1_\mu \phi^* D^{1 \mu} \phi - V(\phi) \nonumber \\
&& + \frac{1}{2} \sum_{i=1}^{N-1} \frac{a z_i^2}{Rg_3^2} 
\frac{\left( A_\mu^{i+1} - A_\mu^i \right)^2}{(a z_i)^2} + \dots
\end{eqnarray}
The dots represent interaction terms in non-Abelian theories, the $D^1$ is a covariant derivative under the first gauge group, and we have suppressed coordinates in the $x^\mu$ direction.

With this discretization, the 2-dimensional gauge theory on each slice, $i$, has a gauge coupling given by:
\begin{equation}
\label{gaugesite}
\frac{1}{g_i^2} = \frac{a z_i^2}{R g_3^2}.
\end{equation}
If we put $\langle \phi \rangle = 0$ so that the gauge symmetry were unbroken by the UV boundary Higgs field then there would be a massless gauge mode comprised of equal parts of the gauge modes from each site.  The corresponding low energy gauge coupling is then approximately given (at tree level) by:
\begin{equation}
\label{gauge2}
\frac{1}{g_2^2} = \sum_{i=1}^N\frac{1}{g_i^2} = \frac{R'^2}{2R g_3^2}.
\label{eqn:IRGaugeCoupling}
\end{equation}
In the presence of the UV brane Higgs mechanism, the low energy gauge coupling comes from considering the overlap of the wavefunctions for the gauge boson and fermions.  However, as the gauge symmetry is restored the above approximation becomes exact.

To find the physical mass spectrum in the gauge sector we need to rescale each gauge boson in order to have the canonical kinetic normalization: $-1/4$.  Doing so gives a gauge mass matrix, $(aR)^2 M_{Gauge} = $
\begin{equation}
\left(\begin{array}{cccccc}
1+v_0^2 & -w & 0 & 0 \\
-w & 2w^2 & -w^3 & 0 & \dots \\
0 & -w^3 & 2w^4 & -w^5 \\
\vdots & & & \ddots & & 0 \\
& & 0 & -w^{2N-5} & w^{2N-4} & -w^{2N-3}\\
& & 0 & 0 & -w^{2N-3} & w^{2N-2}
\end{array}\right), \label{eq:GaugeMassMatrix}
\end{equation}
where $v_0$ is the VEV of the UV boundary Higgs.  In practice it is sufficient to take $v_0=1$ in order to reproduce the mass given by equation~(\ref{eq:LightGaugeBoson}).

Furthermore, the discretization (\ref{warpfactor}) implies $\ln(R'/R) \approx N a$. Thus,  our hierarchy of mass scales, equation~(\ref{eq:MassScaleHierarchy}), can be written, using (\ref{eq:KKGaugeBoson}),(\ref{gaugelight}),(\ref{gaugesite}),(\ref{gauge2}), as:
\begin{equation}
1 \gg a z_i^2 g_i^2 \gg \frac{1}{Na}.
\end{equation}
We see then, that the site gauge couplings must be small in comparison to the local energy scale $1/z_i$.  By choosing to hold $a$ fixed, we satisfy this hierarchy requirement by letting the gauge couplings scale as:
\begin{equation}
\label{couplings}
g_i^2 \sim \frac{1}{z_i^2 \sqrt{N}}
\end{equation}
and taking the large $N$ limit.  With these gauge couplings smaller than other mass scales at site $i$, we do not expect significant corrections to this tree-level relation. 

Finally, we note that the 3d case is different than $AdS_5$; the fact that we can take $N$ large and keep the individual gauge couplings (\ref{couplings}) in  $AdS_3$  small  shouldn't come as a surprise---the 3d bulk theory is superrenormalizable, in contrast to the 5d case where taking large $N$ is ultimately responsible for entering  the strong-coupling domain \cite{Bhattacharya:2005xa}. 


\subsection{Fermions}\label{sec:Fermions}

The presence of an equal number of left and right handed fermions on a lattice means that we must find a way to remove the fermion of one handedness by making it heavy.  In the four dimensional construction of ref.~\cite{Bhattacharya:2005xa}  this could be done indirectly through Majorana mass terms which only give a mass to one Weyl fermion in a Dirac pair. 
 However, in two dimensions Lorentz invariance requires that all mass terms connect a left to a right handed fermion.  While it might be possible to remove one fermion in a Lorentz violating manner, we will choose a different approach: exchanging an unwanted, light, charged fermion for a light neutral fermion. 

We will increase the number of fermions, but make the new fermions neutral under the gauge symmetry.  In the absence of the gauge symmetry breaking at the UV boundary, our low energy spectrum would contain light left and right handed charged Weyl modes, $l_-$ and $l_+$ respectively.  It would also have two neutral Weyl modes, $n_-$ and $n_+$.  We then use the gauge breaking Higgs mechanism from the previous subsection to generate an effective mass term in the low energy theory between the unwanted charged Weyl fermion and one of the neutral fermions:
\begin{equation}
\label{nlcoupling}
y \langle \phi \rangle \bar{l}_+ n_-.
\end{equation}
This mass leaves $l_-$ as the only charged fermion in the low energy spectrum.

All four of these modes, $l_-$, $l_+$, $n_-$, $n_+$, may be realized as domain wall fermions.  In the charged sector, we will maintain consistency with the warped $AdS_3$ background even though this may not always be necessary.  In the neutral sector we will, for simplicity, use flat space domain wall fermions as in Section~\ref{sec:2dDomainWall}.  However, it is important that the fermions in this model are able to reproduce the anomaly for a single, light, left-handed fermion.  Since the fermions are coming from the domain wall, there are an equal and finite number of left and right handed fermions in the measure of the path integral.  The phase of the measure is therefore well defined, so the anomaly will not arise in the usual continuum manner.

The anomaly arises, instead, from symmetry-breaking Majorana masses in the neutral sector and through the couplings (of the form of eqn.~(\ref{nlcoupling})) of neutral and charged modes. We will discuss anomalies and symmetries in more detail in Section~\ref{sec:DomainWallAnomalies}.

We now describe how to obtain the light (before including effects of the UV-brane Higgs) spectrum of charged, $l_-$, $l_+$, and neutral, $n_-$, $n_+$, modes.
 In order to leave a full neutral Weyl spinor $n_\pm$ light we must introduce, in our construction, two neutral Dirac spinors, $n^1_{\pm}$ and $n^2_{\pm}$, so that there will still be a light neutral Weyl spinor after all of the necessary Majorana and Dirac mass terms are included.  In total we will use $3N$ 2-dimensional Dirac spinors for the remainder of this subsection.

We will again use a deconstruction description for the fermions. Alternatively, these fermions may be thought of as ordinary Wilson fermions living on a 2-dimensional lattice in the small lattice spacing limit.  We will use the following fermion basis for expressing the mass matrix:
\begin{eqnarray}
\vec\Psi^T_+ &=& \big(\eta^1_{1+}, \eta^1_{2+}, \dots \eta^1_{N+}, \bar\eta^1_{1+}, 
\bar\eta^1_{2+}, \dots \bar\eta^1_{N+}, \nonumber \\ 
&& \quad \eta^2_{1+}, \eta^2_{2+}, \dots \eta^2_{N+}, \bar\eta^2_{1+}, 
\bar\eta^2_{2+}, \dots \bar\eta^2_{N+}, \nonumber \\ 
&& \quad \psi_{1+}, \psi_{2+}, \dots \psi_{N+}, \bar\psi_{1+}, \bar\psi_{2+},
\dots \bar\psi_{N+} \big),
\end{eqnarray}
and likewise for $\vec\Psi_-$.  In this $6N$ vector of Weyl fermions, the $\eta$'s are all neutral, while the $\psi_i$'s are charged under the $i$th gauge group.  Again, the bar means complex conjugation.  We need to include both the barred and unbarred Weyl fermions in this vector so that we may include the Majorana mass terms.  The mass matrix,
\begin{equation}
\label{PsiMass}
i \vec\Psi^T_+ M \vec\Psi_- ~,
\end{equation}
is almost block diagonal. In addition to the above mass matrix, the complete lagrangian also involves the usual kinetic terms (Wilson, if a 2d lattice description  is used)  for the neutral $\eta^{1,2}_{i \pm}$, $i=1,...,N$ as well as kinetic terms for the charged fermions $\psi_{\pm}^i$ in the warped $AdS_3$ background (see Appendix~\ref{sec:FermionsAdS3} for details).

The mass matrix (\ref{PsiMass}) can be thought of as representing the mass and hopping terms for Dirac fermions in a slice of flat 3-dimensional space---the first $4N \times 4N$ elements---with $\eta_{1\pm}^1$ and $\eta_{1\pm }^2$ being localized at  the left end while the $\eta_{N\pm}^1$ and $\eta_{N\pm}^2$ ``live" at the right end. The right end of the flat space ends at the UV-brane of the $AdS_3$ slice, where $\psi_{1\pm}$ lives. Finally, the $\psi_N$ lives at the IR-brane end of $AdS_3$ (this picture is further visualized by the plots representing the locations of the various modes in Figs.~2, 4). The gauge fields are three-dimensional and propagate in the $AdS_3$ part of the lattice.

We will start by describing the mass and hopping terms    (\ref{PsiMass}) in  its $2N \times 2N$ diagonal blocks,
\begin{equation}
M = \textrm{diag} \left(M_{\eta1}, M_{\eta2}, M_{\psi}\right), \label{eq:BigMassMatrix}
\end{equation}
and add  terms coupling the $\eta$'s to the $\psi$'s at the end of this subsection.  Each of these mass matrices breaks up into $N\times N$ blocks which represent either Dirac or Majorana type masses.  For example, for each $\eta^{1,2}$ and $\psi$, we have a $2N\times2N$ mass matrix of the form:
\begin{equation}
\label{masseta}
M_{\eta} = \left(
\begin{array}{cc}
M_{\eta Majorana} & M_{\eta Dirac} \\
M_{\eta Dirac} & M_{\eta Majorana}
\end{array} \right).
\end{equation}
If we chose $M_{\eta Majorana}=0$ and 
\begin{equation}
a R M_{\eta D} = \left(
\begin{array}{ccccc}
1 & 0 \\
-(1+\epsilon) & 1 & 0 \\
0 & -(1+\epsilon) & \ddots & 0\\
&&& 1 & 0 \\
&&& -(1+\epsilon) & 1
\end{array} \right),
\end{equation}
then we would have an exponentially light, left-handed Weyl mode localized closer to the $N$th slice and an exponentially light right-handed Weyl mode peaked at site $1$.  We will call these modes $n^1_-$ and $n^1_+$, respectively.

If we now add an equal Majorana and Dirac mass term for the light neutral modes, then a Majorana-Weyl spinor will remain massless.  In the low energy theory this mass term has the appearance:
\begin{equation}
\label{MequalsD}
\frac{m}{a}\left(n^1_+n^1_- + \bar n^1_+n^1_- + \textrm{h.c.} \right),
\end{equation}
so that the imaginary part of $n^1_+$, stays massless and $m$ is a number of order one.  (For now, the imaginary part of $n^1_-$ is also massless; note, however, that it is localized near the UV-brane of the warped part of space and  will further get a mass by coupling to the UV-brane localized charged fermions (\ref{eq:ChargedNeutralCoupling}) below.)  This additional mass (\ref{MequalsD}) between the light neutral modes is described  by an entry in the top right corner of both the Dirac and Majorana part of the mass matrix, modifying eqn.~(\ref{masseta}) accordingly:
\begin{eqnarray}
\label{masseta1}
& a R M_{\eta1 M} = \left(
\begin{array}{cccc}
0 & \dots & 0 & 1 \\
&\ddots & 0 & 0 \\
&&& \vdots \\
&&& 0
\end{array}\right), \\
& a R M_{\eta1 D} = \left(
\begin{array}{ccccc}
1 & 0 &&& 1 \\
-(1+\epsilon) & 1 & 0 \\
0 & -(1+\epsilon) & \ddots & 0\\
&&& 1 & 0 \\
&&& -(1+\epsilon) & 1
\end{array} \right). \nonumber
\end{eqnarray}

For the $\eta^2$ masses we can completely repeat this structure with only a change in the sign of the Majorana type mass relative to the Dirac type mass in the top right corner of the matrix (\ref{masseta1}), or equivalently, by adding a mass term to the low energy theory of the form (\ref{MequalsminusD}):
\begin{equation}
\label{MequalsminusD}
\frac{m}{a}\left(n^2_+n^2_- - \bar n^2_+n^2_- + \textrm{h.c.} \right),
\end{equation}
This leaves the real part of $n^2_+$ as an exponentially light mode.  The $n^2_-$ mode is also exponentially light.  However, it is localized near the UV-brane of the warped part of space and will further get a mass by coupling to the UV-brane localized charged fermions, see eqn.~(\ref{eq:ChargedNeutralCoupling}).  Combining the modes of $n^1$ and $n^2$, then, we have a massless Weyl mode localized near site-1 (the leftmost end in Fig.~2), which in the limit $N \to \infty$ exhibits a chiral symmetry:
\begin{equation}
\label{etaplus}
n_+ \equiv \Im (n^1_+) + i \Re (n^2_+) \, \to \, e^{i \alpha} n_+.
\end{equation}

For the charged fermions we make use of the $AdS_3$ background. The continuum action of a Dirac fermion in a slice of $AdS_3$, its
zero mode solutions, and the relevant boundary conditions are given in Appendix~\ref{sec:FermionsAdS3}, where also the discretized version is presented. 
As shown there,  
the Majorana masses for the charged fermions are zero and the Dirac type masses are given by: 

$- 2 a R w M_{D\psi}= $
\begin{eqnarray}
\label{chargedmassmatrix}
= \left(\begin{array}{cccccc}
\alpha & 0 & \dots \\
-\sqrt{w}  w& \alpha w & 0 & \dots \\
0 &   \sqrt{w} w^2 & \alpha w^2 &0&0 \\
0 &  0 & \ddots & 0 &0\\
0 & 0 & \ddots &  \alpha w^{N-2}& 0 \\
0& 0& \dots &-\sqrt{w} w^{N-1}   & \alpha w^{N-1}
\end{array}\right),
\end{eqnarray}
where $\alpha \sim 1$ is defined in Appendix~\ref{sec:FermionsAdS3} and  $w=1/(1+a)$ is the local warp factor. As shown in the appendix, by appropriately choosing $\alpha$, we have a light left-handed mode, $l_-$, peaked at site $N$ (the IR brane).  In addition, the unwanted right-handed companion, $l_+$, is peaked at site $1$ (the UV brane).

\begin{figure}
\includegraphics{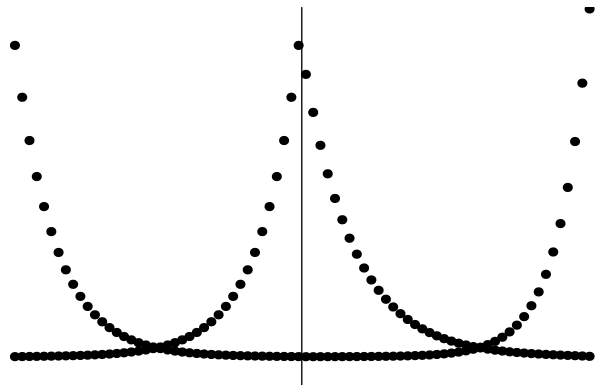}
\caption{The wavefunctions of the exponentially light modes before adding masses which couple the neutral and charged sectors.  The right half is charged with the $l_-$ mode localized on the far right.  The $n_+$ mode is localized on the far left.  The two modes in the middle, $n_-$ and $l_+$ will pick up a mass though the Yukawa coupling with the Higgs on the wall.\label{fig:DomainWallSimple}}
\end{figure}

Finally, to obtain the desired chiral spectrum, we need to couple  the neutral to the charged modes through the UV boundary Higgs.  Before doing that, however, let us take stock of the spectrum.  As shown in figure~\ref{fig:DomainWallSimple}, there are two light Weyl modes: $l_+$ and $l_-$, localized near the UV-brane (middle of figure) and IR-brane (r.h.s.) of the warped part of space. There are also four light Majorana-Weyl modes: $\Im(n^1_-)$, $\Im(n^1_+)$,  $\Re(n^2_-)$, and $\Re(n^2_+)$. The  $\Im(n^1_+)$, $\Re(n^2_+)$---the $n_+$ of (\ref{etaplus}))---are localized at the leftmost end while the   $\Im(n^1_-)$, $\Re(n^2_-)$ are close to the UV-brane.
By showing the wavefunctions, the figure tells us where each mode is localized.  Because the $+$ and $-$ modes are spatially separated we may give a mass to the $l_+$ mode without significant impact on the $l_-$ mode.  This separation is the reason for using the domain wall fermions in the first place.  In the large $N$ limit the $l_-$ mode becomes massless.  In addition, in that limit, the massless neutral modes decouple because their wavefunction has zero overlap with the charged fermions and gauge modes.

The masses which couple neutral and charged modes via the UV brane Higgs show up as off-diagonal entries in the mass matrix from equation~(\ref{eq:BigMassMatrix}).  Instead of writing a large matrix,   their contribution can be written simply in terms of the Lagrangian as:
\begin{equation}
i \phi \left( y_1 \bar\psi_{0+}\eta^1_{N-} + y_2 \bar\psi_{0+}\eta^2_{N-} \right) + \textrm{h.c.},
\label{eq:ChargedNeutralCoupling}
\end{equation}
where $\langle \phi \rangle y_{1,2}$ are of order one.
These masses are sufficient to lift the mass of the $l_+$ mode to the KK scale and to transmit the breaking of the chiral flavor symmetry to reproduce the anomaly as we will discuss below.

\begin{figure}
\includegraphics{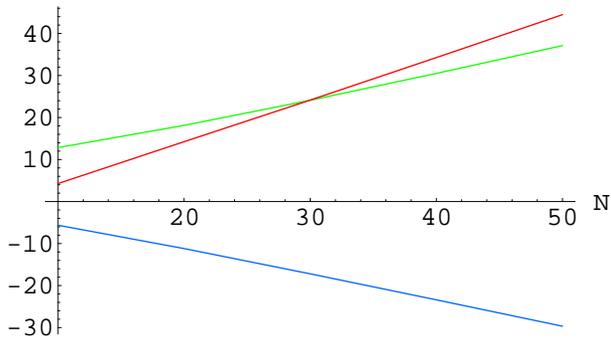}
\caption{(Color online) Mass ratios of the light modes as a function of lattice size, $N$.  The growing lines give $\left({m_{KK}\over m_{A  0}}\right)^2$, the ratio   the KK modes' mass to that of the light gauge boson.  The green (highest intercept) line is for the first gauge KK mode, while the red (middle intercept) line is for the first fermion KK mode.  The falling (negative intercept) line gives the mass of the lightest fermion mode, $m_{f  0}$, by showing $2\ln{ m_{f  0} \over m_{A 0}}$.  Clearly, there is an exponentially light fermion in the spectrum.\label{fig:DomainWallSpectrum}}
\end{figure}

The qualitative expectations for the mass spectrum---based on known domain-wall fermion spectra and on fermion spectra in $AdS$  backgrounds, see Appendix~\ref{sec:FermionsAdS3}---can be further substantiated by numerically solving for the eigenvalues and eigenvectors of the mass matrices for various values of $N$.
Figure~\ref{fig:DomainWallSpectrum} shows the mass spectrum resulting from the diagonalization of the mass matrices in equations~(\ref{eq:GaugeMassMatrix}),~(\ref{eq:BigMassMatrix}), and~(\ref{eq:ChargedNeutralCoupling}).  This figure includes the results of all mass terms discussed in this section.  The mass ratios of the first KK modes to the lightest gauge mode, $m^2_{A1}/m^2_{A0}$ and $m^2_{f1}/m^2_{A0}$, are plotted as a function of $N$.  The KK modes are getting heavier than the light gauge mode as expected.  The log of the ratio of the light fermion to the light gauge mode, $2\log(m_{f0}/m_{A0})$, is also plotted and shows that one light charged fermion remains.

\subsection{Anomalies}\label{sec:DomainWallAnomalies}
The condition for a 2-dimensional $U(1)$ gauge theory to be free of gauge anomalies is that the sum of the squares of the charges of the left and right handed modes must cancel:
\begin{equation}
\sum_i q^2_{i,left} - \sum_j q^2_{j,right} = 0.
\end{equation}
An example of such a theory which we will use here is the ``345" theory where there are left-handed fermions of charge $3$ and $4$ as well as right handed fermions of charge $5$:
\begin{equation}
3_-, \, 4_-, \, 5_+.
\end{equation}
Before adding gauge breaking mass terms, our warped domain wall construction necessarily contains the mirror fermions as well, $3_+, \, 4_+, \, 5_-$.  As mentioned above, the Lorentz structure requires that left and right handed modes be lifted in pairs and so at least one neutral mode is needed in order to provide enough mass terms to remove the unwanted charged modes.

If only Dirac masses are contained in the theory, then one global fermion number $U(1)$ symmetry will remain.  A current of charged fermions may end up in neutral modes, with the gauge charge absorbed by the Higgs field.  However, the target continuum theory violates fermion number and the 't Hooft operator has the (schematic) form:
\begin{equation}
\label{thooftvertex}
(3_-)^3\; \partial_+ (4_-)^4 \; (\bar{5}_+)^5~,
\end{equation}
where $4_-$ denotes a Weyl fermion field representing a left-handed Weyl fermion of $U(1)$ charge 4, etc.  Therefore Majorana masses are needed in order to introduce a violation of fermion number into the lattice (this has already been noted in \cite{Golterman:2002ns}).  

\begin{figure*}
\includegraphics[width=14cm]{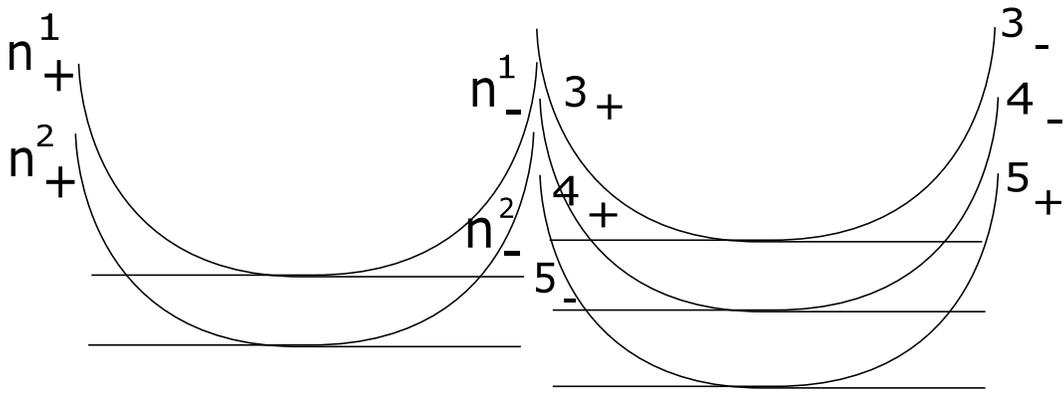}
\caption{A schematic representation of the location of all (exponentially light) modes which are needed for the 345 theory to have the correct anomaly properties.  The right half is gauged; the left is neutral.  After introducing the UV-brane mass terms from equations~(\ref{MequalsD},\ref{MequalsminusD},\ref{eq:ChargedNeutralCoupling}), the only remaining light modes will be the $3_-$, $4_-$, and $5_+$ from the righthand side as well as one Weyl combination of the $n^1_+$ and $n^2_+$ from the lefthand side. \label{fig:DomainWallFermions}}
\end{figure*}

To describe them, recall that,  as discussed in the previous subsection, we introduced two neutral Dirac modes $\eta^{1,2}$, which led, before adding any of the mass terms (\ref{MequalsD},\ref{MequalsminusD},\ref{eq:ChargedNeutralCoupling}),  to four Weyl fermions---the left-handed ones ($n_-^{1,2}$) localized near the UV-brane and the right-handed  ones ($n_+^{1,2}$), localized at the far left in the ``flat slice bulk''  The wavefunctions of the lightest modes  of $\eta^{1,2}$ as well as those of the ``$345$" Dirac fields are shown in figure~\ref{fig:DomainWallFermions}.

We now add the equal or opposite strength Majorana and Dirac masses, (\ref{MequalsD}, \ref{MequalsminusD}), to the neutral modes, so that one massless neutral Dirac mode remains---the $n_+$ at the far left end and $n_-$ near the UV-brane.  Finally, we add the mass terms   (\ref{eq:ChargedNeutralCoupling}), coupling  the unwanted charged mirror modes and the $n_-$ mode.  In addition, we add Majorana mass terms of the form $\eta_{N \; -}^1 \psi_{1\; +}^3$ (and similar for all unwanted charged modes ($\psi_{1 \;+}^3$,  $\psi_{1 \;+}^4$, $\psi_{1 \;-}^5$, including appropriate powers of the higgs field) violating the fermion number symmetry.
Note that there are an equal number of left and right handed fields to which we are giving a mass.

With all of the masses discussed above, the only remaining exact symmetry in the theory is the global part of the gauge symmetry, the ``345" symmetry.  However, the 't Hooft operator preserves also another global symmetry, 133, where the $4_-$ and $5_+$ transform with three times the phase of the $3_-$.  We speculate that either this 133 symmetry will emerge in the IR, or else we have found a theory which preserves no symmetry beyond the gauged 345,  which can happen  if 133-violating operators (e.g. four-fermi operators in 2 dimensions) remain relevant in the IR.  Later on, we will present a ``one-site'' model with Ginsparg-Wilson fermions, where both the 345 and 133 symmetries are exact symmetries of the partition function while the fermion number has the correct anomaly.

\subsection{Scalar Couplings}\label{sec:Scalars}
While the gauge couplings appear to be perturbative, we verify here that the longitudinal mode of the gauge boson is not strongly coupled to the fermion wavefunctions as it is in the $AdS_5$ case.  We begin by writing in the gauge terms and then calculating the relevant Yukawa coupling.

We will do this calculation in the 3-d continuum language in $AdS$ where an analytic expression for the gauge boson wavefunction may be found.  We take the lightest longitudinal gauge mode to be:
\begin{equation}
A_\mu = f_0(z) \partial_\mu \varphi(x).
\end{equation}
By using the previously found wavefunction for arbitrary mass, equation~(\ref{eq:besselWaveFunctions}), and considering the kinetic terms for the gauge field we can find the proper normalization:
\begin{equation}
f_0(z) = \sqrt{R} \ln (z/R). \label{eqn:zeroModeWavefunction}
\end{equation}
To see how this longitudinal mode couples to the fermions we look at the gauge terms in the fermion kinetic term:
\begin{equation}
\int d^2x dz \left(\frac{R}{z}\right)^2 (-2i) (-ig_3) 
\left( \bar\psi_- A_+ \psi_- + \bar\psi_+ A_- \psi_+ \right),
\end{equation}
where we are working in the $A_3=0$ gauge.  In terms of the longitudinal mode after an integration by parts we have
\begin{equation}
2 g_3 \int d^2x dz \left(\frac{R}{z}\right)^2 f_0(z) \varphi 
\left[ \partial_+ (\bar\psi_-\psi_-) + \partial_- (\bar\psi_+\psi_+) \right].
\end{equation}
After expanding these derivatives we have four terms involving derivatives on the fermions.  By making use of the equations of motion we can turn these four terms into expressions involving the fermion masses, the $\partial_z$, and the gauge modes.  The terms involving the bulk mass and gauge modes cancel, leaving us with
\begin{eqnarray}
&& g_3 \int d^2x dz \left(\frac{R}{z}\right)^2 f_0(z) \varphi \Big[ -\partial_z \bar\psi_+ \psi_-
- \bar\psi_-\partial_z \psi_+ \nonumber \\
&& \:\; -\partial_z\bar\psi_-\psi_+ - \bar\psi_+\partial_z \psi_-
+ \frac{2}{z}(\bar\psi_+\psi_- + \bar\psi_-\psi_+) \Big]~.
\end{eqnarray}
We can now perform an integration by parts in $z$ on two of these terms (for example the first two).  Most terms cancel (including the $2/z$ terms) leaving us with:
\begin{equation}
g_3 \int d^2x dz \left(\frac{R}{z}\right)^2 \partial_z f_0(z) \varphi(x) (\bar\psi_+\psi_- + \bar\psi_-\psi_+).
\end{equation}
The fermion kinetic term has the same leading factor of $(R/z)^2$ and so we should rescale the fields to get canonical kinetic terms.  This leaves us with a Yukawa coupling between the longitudinal mode, $\varphi$, and the fermions:
\begin{equation}
y(z) = g_3 \partial_z f_0 = \frac{g_3 \sqrt{R}}{z}.
\end{equation}
From the scaling requirements  in section~\ref{sec:GaugeFields} we found that we must have $g_3^2 R \ll 1$.  This means that our Yukawa coupling is smaller than the local scale $1/z$ which sets the fermion masses at that location.  In addition, this Yukawa coupling has units of mass as we expect for a coupling between a 2-d scalar and two 2-d fermions.

To see that this Yukawa really is perturbative we can estimate the size of loop corrections to the fermion mass and kinetic terms.  Consider one loop contributions to the fermion two point function which involve one scalar and one fermion in the loop.  The external fermion legs are in the site basis (at sites $i$ and $j$), but the propagator in the loop must be in the mass eigenstate basis.  This gives a contribution:
\begin{equation}
y_i y_j \sum_k \alpha_{ik} \alpha_{jk} \int d^2p 
\frac{\gamma^\mu p_\mu - m_\psi^{(k)}}{p^2 - m_\psi^{(k)2} + i\epsilon}
\frac{1}{(q-p)^2 - m_\varphi^2+ i\epsilon},
\end{equation}
where $\alpha$ relates the site and KK bases:
\begin{equation}
\Psi_i = \sum_k \alpha_{ik} \Psi^{(k)}.
\end{equation}
After combining the denominators and shifting the momentum we see that on dimensional grounds the result scales as
\begin{equation}
\sum_k \alpha_{ik} \alpha_{jk} \frac{y_i y_j}{m_\psi^{(k)2}} 
\left(\gamma^\mu q_\mu - m_\psi^{(k)}\right). \label{eq:KKSum}
\end{equation}
First, let us heuristically argue that this is small and then compute this matrix using our numerical solutions for the wavefunctions.  The KK modes tend to be localized in that part of the space which corresponds to their mass: $z \sim 1/m$.  If the modes were exactly localized then the matrix $\alpha$ would be diagonal.  Furthermore $\alpha$ is unitary and the masses scale like $m^{(k)} \sim 1/(az_{k})$ for most of the KK modes.  In that case we have
\begin{equation}
R g_3^2 a^2 \delta_{ij} \left(\gamma^\mu q_\mu - m_\psi^{(i)}\right).
\end{equation}
This is clearly a small correction to the action at each site.

In fact the matrix relating site and KK bases is not diagonal, but we can find a numerical solution for $\alpha_{ij}$ and the KK masses.  We may then calculate this sum over KK modes in equation~(\ref{eq:KKSum}).  (This sum includes the charged mode, $l_+$, which became heavy along with the neutral mode, $n_-$).  Doing this shows that the results are in fact a small number (of order $a^2$) times our leading small factor of $R g_3^2$.  Note that the exchange of the fermion zero modes has been ignored because it is IR divergent.  However, that IR divergence is present in the target theory as well, so it is to be expected (see Appendix B for a calculation of the domain wall beta function in the deconstucted version of the theory).


\section{A 1-site Construction}\label{sec:2Site}

In this section, we present a one-site model using Ginsparg-Wilson (GW) fermions which has precisely the light field spectrum discussed at length above. It also exhibits  {\it exactly} the right set of symmetries and anomalies to be a candidate for a lattice formulation of the 345 theory. The advantage of this formulation is that its chiral symmetries---which are only expected to emerge at large $N$ in the warped domain wall model---are exact symmetries of the lattice theory. Thus one  can study their consequences, including the associated exact (anomalous or non-anomalous) lattice Ward identities. 

Furthermore, a strong-coupling analysis at the end of this section indicates that the spectrum of this theory is chiral and that this proposal may be a road to constructing the fermion measure for chiral gauge theories with GW fermions starting from a vectorlike theory, where the measure is well defined.

In essence, the idea is to consider a ``one-site limit" of our construction of Section III, using 2-dimensional GW fermions in order to implement exact lattice chiral symmetries.  Schematically, the field content and couplings of the model are represented in Fig.~5. There are, for the 345 $U(1)$ theory,  three 2-dimensional Dirac fermions, $\Psi_{3}$, $\Psi_4$, $\Psi_5$, charged under the $U(1)$ gauge group with charges $3,4,5$, respectively. There is also a neutral Dirac fermion,  $\Psi_0$. 
\begin{figure}
\includegraphics[width=7cm]{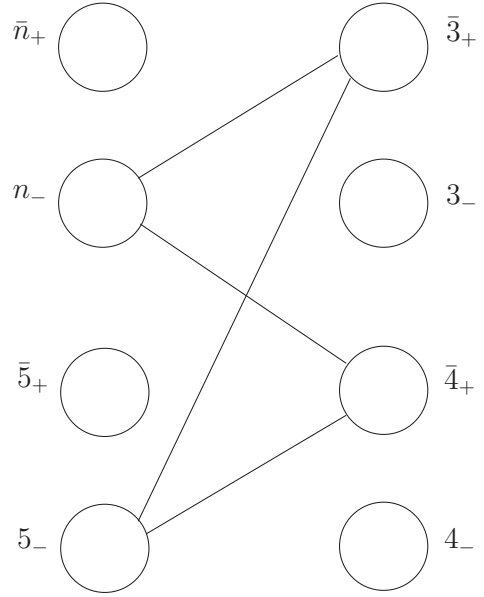}
\caption{The 2-dimensional model using GW fermions.  The lines represent arbitrary ${\cal O}(1/a)$ masses of both Dirac and Majorana type.  Due to the chiral symmetry present in the GW formulation, each fermion is exactly massless before an explicit mass term is added.  Therefore, four modes remain massless: $n_+$, $3_-$, $5_+$, and $4_-$.\label{fig:GWFermions}}
\end{figure}

The fermion fields live on the sites, labeled by $\{x\}$, of a two dimensional lattice and their lattice action consists of kinetic terms:
\begin{eqnarray}
\label{kinetic}
S_{kin} =\sum_{q = 0,3,4,5} \sum_{x,y} \bar\Psi_{q}(x) D_q(x,y) \Psi_q(y)~,
\end{eqnarray}
where $D_q$ is the GW operator for a fermion of charge $q$, obeying the GW relation (for a review of the GW relation and exact chiral symmetry on the lattice, see, for example \cite{Luscher:2000hn} and references therein):
\begin{equation}
\label{GW}
\{D_q,  \gamma_5 \} =   D_q \gamma_5 D_q ~.
\end{equation}
Here $\gamma_5$ is the appropriate matrix in 2d and the lattice spacing  has been set (from now on) to unity.
The     lattice action (\ref{kinetic}) has a large number of exact global symmetries: 
\begin{eqnarray}
\label{kineticsymmetries}
\prod_{q = 0,3,4,5}  U(1)_{q, -} \times U(1)_{q,+}~,
\end{eqnarray}
where $U(1)_{q,\pm}$ acts only on the Dirac fermion field of charge $q$ as follows:
\begin{eqnarray}
\label{kineticsymmetries2}
\Psi_q &\rightarrow& e^{i \alpha_{q, \pm} P_\pm} \Psi_q \nonumber \\
\bar\Psi_q &\rightarrow& \bar\Psi_q e^ {-i\alpha_{q, \pm} \hat{P}_\mp}~,
\end{eqnarray}
where $P_\pm = (1 \pm \gamma_5)/2$ and $\hat{P}_\pm = (1 \pm \hat{\gamma}_5)/2$ with
 $\hat{\gamma}_5 \equiv (1 - D) \gamma_5$ ($\hat{\gamma}_5^2 = 1$ follows from the GW relation (\ref{GW}); also note that  $\Psi_q$ and $\bar\Psi_q$ transform differently, which is perfectly natural in Euclidean space). The projector used for every $\Psi_q$ involves the appropriate GW operator $D_q$.

That the symmetries in equation~(\ref{kineticsymmetries2}) are all exact
follows from $\hat{P}_\mp D = D P_\pm$---yet another consequence of the GW relation (\ref{GW}). Furthermore, the measure of integration is not invariant under any individual $U(1)_{q, +}$ or $U(1)_{q, -}$. Instead, under a   $U(1)_{q, \pm}$ transformation (\ref{kineticsymmetries2}) with parameter $\alpha_{q, \pm}$, the measure  changes:
\begin{eqnarray}
\label{changeofmeasure} 
& &U(1)_{q, \pm}: \prod_{r = 0,3,4,5} d\bar\Psi_r d\Psi_r  \rightarrow  \nonumber \\
&  \rightarrow &\prod_{r = 0,3,4,5} d\bar\Psi_r d\Psi_r  \left[1 -i \alpha_{q, \pm} {\rm Tr} \left( P_\pm - \hat{P}_{\mp}   \right)\right]    \\
&=& \prod_{r = 0,3,4,5} d\bar\Psi_r d\Psi_r  \left[1 \pm i \alpha_{q, \pm} {\rm Tr} \left(\gamma_5 - {1 \over 2} D_q  \gamma_5\right) \right]~. \nonumber
\end{eqnarray}
Eqn.~(\ref{changeofmeasure}) implies that 
for vectorlike symmetries $U(1)_{qV}$ ($\alpha_{q,+} = \alpha_{q, -}$),  there is no Jacobian and thus they are true symmetries of the theory. On the other hand \cite{Hasenfratz:1998ri,Fujikawa:1998if}, since ${\rm Tr} (\gamma_5 - {1 \over 2} D_q  \gamma_5) = n_+^0 - n_-^0$ (the difference between the number of  left- and right-handed zero modes of $D_q$), the continuum violation of charge for anomalous symmetries is reproduced  by the nonzero Jacobian.

To construct our candidate ``345" chiral lattice theory, we introduce a unitary higgs field, $\phi(x)$,  living on the lattice sites (we assume that the issues with building a UV-completion, or  the necessity thereof, see \cite{Fradkin:1978dv}, for the unitary Higgs field are independent of the problem of chirality on the lattice).   We will  use $\phi(x)$ to write all possible Dirac and Majorana  mass terms that violate all symmetries (\ref{kineticsymmetries}) of the kinetic term  (\ref{kinetic}) except:
\begin{equation}
\label{masssymmetries}
U(1)_{3,-}\times U(1)_{4,-} \times U(1)_{5,+} \times U(1)_{0, +}~.
\end{equation}
 The explicit form of the  mass matrix is described in what follows, eqn.~(\ref{DMmass}), and is also schematically indicated on fig.~5. The lattice path integral measure is not invariant under all four $U(1)$ symmetries (\ref{masssymmetries}) of action.  It only respects  three linear combinations: the $U(1)_{345}$  and the $U(1)_{133}$ chiral symmetries---linear combinations of  $U(1)_{3,-} \times U(1)_{4,-} \times U(1)_{5,+} $ with coefficients $345$ and $133$, respectively---and the $U(1)_{0,+}$, which acts only on the $n_+ \equiv P_+ \Psi_{0}$ neutral field, whose dynamics is expected to decouple from the physics of the charged sector.

The $345$ and $133$ $U(1)$'s are exact global symmetries of the partition function. On the other hand, the third linear combination of the first three $U(1)$'s in equation~(\ref{masssymmetries})---the fermion number symmetry of the light fields, which can be taken to be the ``$111$'' symmetry---has an anomaly exactly reproduced by the Jacobian, eqn.~(\ref{changeofmeasure}), of the corresponding transformation  of the measure; see  \cite{Hasenfratz:1998ri}, \cite{Fujikawa:1998if}, and references in \cite{Luscher:2000hn}. Thus, the lattice theory obeys exact Ward identities, including the anomalous ones. For example, using (\ref{changeofmeasure}) one finds  that  the 111 transform of an operator $\cal{O}$ obeys the exact lattice Ward identity: 
\begin{eqnarray}
\label{WI}
\langle \delta_{\alpha_{111}}  {\cal{O}} \rangle =
i\; {\alpha \over 2} \; \langle  {\cal{O}} \; {\rm Tr}\left[ \gamma_5 (D_3 + D_4 - D_5)  \right]  \rangle~.
\end{eqnarray}
The continuum limit expansion  Tr$\gamma_5 D_q \sim  \int d^2 x \; \epsilon^{\mu \nu} F_{\mu \nu}$  \cite{Fujikawa:1998if} implies that the anomalous Ward identity (\ref{WI}) has 
 a continuum limit exactly as expected.

To ensure that the dynamics of this theory reproduces that of the desired unbroken chiral gauge theory, we next focus our attention on the coupling of the Higgs field to the fermions, as well as on its kinetic term (i.e., the mass term for the gauge field). In particular, we will study the possible existence of the strong-Yukawa-coupling symmetric phase (recall again the strong coupling analysis of \cite{Golterman:1994at} which showed that in the waveguide model the spectrum in this phase was vectorlike). Remarkably, as we find below,  to leading order in the strong Yukawa coupling expansion and small gauge coupling---precisely the regime where the waveguide idea broke down---there appear no new massless modes and the spectrum of the unbroken gauge theory is now chiral.

We begin by writing the most general mass matrix which breaks the symmetries of equation~(\ref{kineticsymmetries}) to the four chiral $U(1)$'s of (\ref{masssymmetries}). To this end, we relate the Dirac fields $\Psi_q$ to their  chiral components: $\Psi_{q, \pm} \equiv P_{\pm} \Psi_q$, $\bar\Psi_{q,\pm} \equiv \bar\Psi_{q} \hat{P}_{\mp}$; note that the definition of the   $\bar\Psi_\pm$ chiral  modes is now both momentum and gauge-background dependent. 
We then write down the most general Dirac and Majorana couplings---giving mass  of order the inverse cutoff---to the fields:
\begin{eqnarray}
X_+ &=& (\Psi_{3,+}^T \: \bar\Psi_{3,+} \: \Psi_{4,+}^T \: \bar\Psi_{4,+}) \nonumber \\
Y_- &=& \left(\begin{array}{c} \Psi_{5,-} \cr \bar\Psi_{5,-}^T \cr \Psi_{0,-}\cr \bar\Psi_{0,-}^T\end{array}\right)~,
\end{eqnarray}
where $T$ denotes transposition (we treat unbarred Dirac spinors as columns and barred ones as rows) of the form:
\begin{eqnarray}
\label{DMmass}
S_{mass} = \lambda \sum_{x} X_+(x) M Y_{-}(x)~.
\end{eqnarray}
The structure of $S_{mass}$  is evident from fig.~5, where both Dirac and Majorana masses are  to be included for the connected fields. We note that if Majorana masses are omitted, there will be extra unbroken chiral symmetries and unlifted zero modes in an instanton background, resulting in failure to reproduce  the 't Hooft vertex (\ref{thooftvertex}); moreover, consistent with the symmetry argument, a careful analysis shows that with  Dirac masses only, the mass matrix (\ref{DMmass}) has a zero eigenvalue at the end of the Brillouin zone; details will be given in a future publication.

Instead of writing explicitly the entire matrix $M$, we give an example of  a Dirac mass term: $\bar\Psi_{0,-} (\phi^{*})^3 \Psi_{3,+}  + \bar\Psi_{3,+} \phi^3\Psi_{0,-} $,  and of  a Majorana mass of the form: $\bar\Psi_{5,-} \gamma_2 \phi^8 (\bar\Psi_{3,+})^T - \Psi_{3,+}^T  \gamma_2 (\phi^{*})^8 \Psi_{5,-}$. Here $\gamma_2$ is the (hermitean) 2d gamma matrix that appears when Majorana masses are written using Dirac spinors, while $\phi$ is the unitary Higgs field. Thus, the explicit form of $M$ in (\ref{DMmass}) contains appropriate powers of $\phi$ and $\gamma_2$-insertions.
The general mass matrix (\ref{DMmass}) violates all $U(1)$ symmetries from (\ref{kineticsymmetries}) and preserves the desired $U(1)_{3,-}\times U(1)_{4,-} \times U(1)_{5,+} \times U(1)_{0, +}$ symmetry (\ref{masssymmetries}). 

The total action of our lattice model is, finally: 
\begin{eqnarray}
\label{totalL}
S& =& S_{Wilson} + S_{kin} + S_{mass} \\
&+& {\kappa \over 2}\; \sum_{x} \sum\limits_{\hat{\mu}} \left[ 2 - \left(\phi(x)^* U(x, \hat\mu) \phi(x+\hat\mu) + {\rm h.c.}\right) \right]~, \nonumber~
\end{eqnarray}
where $S_{kin}$ is defined in (\ref{kinetic}), $S_{mass}$---in (\ref{DMmass}),  the last term is the kinetic term for the charge-1 unitary lattice Higgs field $\phi$, and $S_{Wilson}$ is the usual plaquette action for the link variables $U(x, \hat\mu)$ (appropriately modified to restrict the gauge field path integral to  admissible gauge field backgrounds, see \cite{Luscher:2000hn}).

In the broken phase, when $\langle \phi \rangle \ne 0$, we already   analyzed the fermion  spectrum  and found that  there are four light modes: the charged $\Psi_{3,-}$,  $\Psi_{4,-}$,  $\Psi_{5,+}$,  and the neutral $\Psi_{0,+}$. The dynamical issue that needs to be addressed is the existence of an unbroken phase where $\langle \phi \rangle = 0$, such that the gauge boson is ``massless." The gauge symmetry can thus be thought of as ``emerging" in the IR  \cite{Forster:1980dg}. The essential idea behind the  ``FNN mechanism"  \cite{Forster:1980dg} is that integrating out the fluctuations of the unitary Higgs field, whose  correlation length in the symmetric phase is a few lattice spacings, results in   renormalization of the gauge coupling plus a tower of higher-dimensional gauge invariant local operators which are irrelevant for the long-distance physics of the gauge theory. The fact that lattice-Higgs models exhibit such behavior  is well known;  for the 2-dimensional case, see also \cite{Einhorn:1977jm},  and \cite{Fradkin:1978dv} for a general analysis in various dimensions. Previous discussions of the use of this mechanism to the lattice definition of chiral gauge theories are given in refs.~\cite{Smit:1987en, Parisi:1992qz}.

In our case, an important requirement further to the ``restoration" of the gauge symmetry at distances larger than a few lattice spacings should be that   there are no new massless fermions in addition to the desired massless chiral spectrum.

To study  the continuum limit in the asymptotically free theory it is sufficient to begin at leading order in the $g \rightarrow 0$ expansion; in fact, apart from a few comments, here we will confine our analysis to this limit. This freezes the gauge degrees of freedom to $U=1$. The resulting theory is a unitary Higgs-Yukawa model whose phase structure can be studied in various limits.  We are interested in the symmetric phase of the lattice $O(2)$ model and will  take $\kappa < \kappa_c$ (simple random-walk intuition  leads  to the estimate   $\kappa_c^{-1} \geq 2d$ in $d$ dimensions on  a hypercubic lattice  \cite{Forster:1980dg}) while also taking the $\lambda \rightarrow \infty$ limit. Recall that this was precisely the limit where the strong coupling analysis of the waveguide showed that  new massless fermions were appearing at the waveguide boundary, see our discussion at the end of Section II and ref.~\cite{Golterman:1994at}.

Since $d \Psi= d \Psi_+ d \Psi_-$, the lattice partition function factorizes, in a trivial gauge background, into a product
$Z = Z_{light} \times Z_{mirror}$:
\begin{eqnarray}
\label{Zmirror}
Z_{light} &=& \int \prod_{x} d \Psi_{3,-} d \Psi_{4,-} d \Psi_{5,+}d \Psi_{0,+}  e^{- S_{kin}(\Psi^{light})} \nonumber \\
 Z_{mirror}&=& \int \prod_{x} d \Psi_{3,+} d \Psi_{4,+} d \Psi_{5,-}d \Psi_{0,-} d \phi \\
 \times  &&  e^{- S_{kin}^{mirror}(\Psi^{mirror}) - S_\kappa(\phi) - S_{mass}(\Psi^{mirror}) } ~. \nonumber
\end{eqnarray}
For conciseness, we  omitted the conjugate fields in the measure and denoted collectively by $\Psi_{light}$ the fields $\Psi_{3,-}, \Psi_{4,-}, \Psi_{5,+}$, $\Psi_{0,+}$, and by $\Psi_{mirror}$ the heavy charged mirrors $\Psi_{3,+}, \Psi_{4,+}, \Psi_{5,-}$, and the neutral $\Psi_{0,-}$. The mass term is given by equation~(\ref{DMmass}) and  the kinetic term for $\phi$ by~(\ref{totalL}). 

The most important point is the splitting of the kinetic terms (\ref{kinetic}) into  light and mirror modes in (\ref{Zmirror}). This  follows from the identity (note that it also holds in an  arbitrary gauge background):
\begin{eqnarray}
\label{kinetic2}
\bar\Psi_{q}  D_q  \Psi_q = \bar\Psi_{q, +} D_q \Psi_{q, +} + \bar\Psi_{q,-} D_q \Psi_{q,-}~,
\end{eqnarray}
where the   cross terms vanish due to the GW relation (\ref{GW}). Thus the mirror and light partition functions factorize at $g=0$; recall our discussion of Section II  showing that the lack of factorization in the kinetic terms  was the cause of failure of the waveguide.  Of course, for $g \ne 0$ the factorization of the measure depends on the gauge field (see discussion in the following paragraphs), but we are only interested in the spectrum of the Yukawa-Higgs model at this point.  We stress that the GW relation was crucial in order for (\ref{kinetic2}) to hold; we know of no other way to achieve (\ref{kinetic2}) and hence the factorization (\ref{Zmirror}) on the lattice.  

Finally, let us study $Z_{mirror}$ and its effect on the light modes, in the $\lambda \rightarrow \infty$ and $\kappa < \kappa_c$ limit. Of particular concern is the possible appearance of extra massless states and the associated vanishing of the mirror determinant. 
To this end, we redefine the mirror fermion fields in (\ref{Zmirror}) ($\Psi_{3,+}, \Psi_{4,+}, \Psi_{5,-}$, and the singlet $\Psi_{0,-}$) by $1/\sqrt{\lambda}$. This multiplies their kinetic terms by $1/\lambda$. Thus, as $\lambda \rightarrow \infty$, the mirror fields kinetic terms vanish, and the mirror  action consists solely of a mass term given by (\ref{DMmass}) with $\lambda = 1$. We can now perform the  integral over the mirror fermions in $Z_{mirror}$, leading to a factor of det$M$---by construction manifestly nonzero and $\phi$-independent. Hence, to this order of the strong coupling expansion, there are no new massless states. 

Admittedly, the argument of the previous paragraph is  oversimplified. The true story is more complicated, due to the fact that the $\bar\Psi_\pm$ chiral components are somewhat smeared due to the nonlocality of the chiral projectors, and will be explained in \cite{GP2006}. Nevertheless, the results there indicate that  the scalar dynamics is not significantly affected by the fermions' quantum fluctuations, with the conclusion that the ``FNN mechanism" continues to apply, and  that there are no light mirror modes.

Ideally, turning on a small gauge coupling will not cause a dramatic rearrangement of the spectrum. While the $g \ne 0$ case clearly deserves further detailed study, we expect that the effect of the mirror fermions on the gauge field and the light chiral fermions is parametrically suppressed  by $1/\lambda$ (the one notable  exception should occur if  the massless fermion spectrum is  anomalous, when   the gauge coupling  is turned on,  a mass  for the gauge boson  is generated  \cite{Jackiw:1984zi, Preskill:1990fr}, with the details controlled by the ultraviolet physics).  
To argue for this, we  note that the factorization of the partition function into mirror and light, eqn.~(\ref{Zmirror}), due to (\ref{kinetic2}), occurs also in fixed nontrivial gauge backgrounds  (details, including the factorization of the fermion measure, are under investigation and  will be given elsewhere).  
The integral over the mirror fermions can now be performed as in the $U=1$ case above, by noting that the  $U \ne 1$ gauge field background  interacts with the mirror fermions only through their kinetic terms. Thus, one expects that all effects of the mirror fermions on the gauge field effective action are local and of order $1/\lambda$. Taking into account the kinetic terms of $\phi$ and the mirror fermions in a strong coupling expansion leads to corrections to $\kappa$:  $\kappa \rightarrow \kappa + {\cal{O}}(1/\lambda)$, as well as to other   ${\cal{O}}(1/\lambda)$ terms, like $\sum_{x} (\phi(x))^3 (\phi(x+\mu)^*)^3$, etc., including higher powers of $\phi$.  A detailed study of the phase diagram away from the $1/\lambda$ expansion  is beyond our scope here; we stress again that this stage of our analysis---the $g=0$ analysis of the Yukawa-Higgs model---was precisely where the waveguide model failed
\cite{Golterman:1994at} to reproduce the chiral gauge theory spectrum.

We should also note that nothing (except for the need, coming from  2 dimensional Lorentz invariance, to introduce the spectator neutral fermions) about the proposal considered in this section is  intrinsically 2-dimensional. In fact, all the steps and relevant properties, including the factorization (\ref{kinetic2}) of the Ginsparg-Wilson fermion kinetic terms and the  existence of a ``high-temperature" disordered phase of the compact  Higgs variables, hold in a four-dimensional theory as well,  particularly in the abelian case considered here. A  more detailed study of a similar construction of nonabelian chiral theories will be given elsewhere.

Finally, we reiterate why we think that this ``one-site" proposal is of interest. It is 
{\it a.)} a full lattice proposal (not deconstructed---all dimensions are latticized) of a local action and measure for a chiral gauge theory, {\it b.)} the realization of both the anomalous and anomaly-free global symmetries is exactly as in the target continuum theory, and, {\it c.)} we gave plausibility arguments why the FNN mechanism may work and the breaking of gauge symmetry be irrelevant in the infrared.

 While we have not proven that the proposal results in a chiral lattice gauge theory,
we believe that the three points above  warrant its presentation and further study. 
Clearly, the study of the $g=0$ dynamics currently underway \cite{GP2006} has to be followed by a detailed study of the $g\ne0$ case, both in perturbation theory and nonperturbatively, and by a convincing demonstration that  an unbroken lattice gauge theory with a chiral spectrum of fermions has been constructed.

\section{Summary, Relation Between the Two Models, and outlook}

Let us first summarize the main results of this paper. 
\begin{enumerate}
\item{
We began by a study of  the earlier proposal of  the warped domain wall model \cite{Bhattacharya:2005xa}. 
Motivated by the strong-coupling issues  encountered by  this proposal in 4-dimensions \cite{Bhattacharya:2005xa}, we turned to the simpler 2-dimensional case, where 
the target 2-dimensional chiral theory is the IR limit of a 3-dimensional theory in a slice of $AdS_3$. We studied in  detail the spectrum and perturbative expansion of the deconstructed version of the theory (it is expected that this analysis is adequate also for small enough lattice spacing in the two  dimensions of the target theory).

 We showed through a perturbative analysis that in the $N \to \infty$ limit  the IR theory has massless gauge bosons and a chiral spectrum of massless fermions in the weak coupling regime. We found no strong coupling of the goldstone mode to the fermions, in contrast to the 4-dimensional case.  Thus, while  our 2-dimensional study has nothing to say about the viability of warped domain walls in the physically interesteing case of 4 dimensiona, it indicates that  this proposal is still of interest and worthy of further study. We believe that it is  a useful first step towards the full lattice study of this proposal (which still awaits implementation). }
\item{
Next, we proposed  a purely 2-dimensional lattice theory, a ``one-site model." It uses the Ginsparg-Wilson mechanism of   exact realization of  chiral global symmetries at finite lattice spacing. The model has   modified, momentum and gauge-background dependent, chiral symmetries, which  reduce  to the usual continuum chiral symmetries  for the low-lying modes. The exact chiral symmetry also ensures that the Ward identities at finite lattice spacing are the ones of the continuum theory.  We argued, in a preliminary  strong Yukawa coupling analysis, for the existence of an unbroken phase with a chiral spectrum of fermions (at $g=0$), in contrast to the analogous phase of the waveguide model. Admittedly, a more detailed analysis is needed; in this regard, we note that the forthcoming results of \cite{GP2006} offer a strong indication that the plausibility arguments given in Section IV indeed hold. 
}
\end{enumerate}
The common theme of the two proposals is that the chiral spectrum is obtained after a particular limit of a vectorlike theory is taken, which decouples the mirrors while keeping the gauge boson massless. Thus, both proposals are similar to the "waveguide" models. 

The first proposal uses warping and localization to address the weak-coupling problems of the waveguide models, discussed in Section II and ref.~\cite{Golterman:1993th}) where the Higgs-Yukawa sector of the theory is in the broken phase. On the other hand, the "one-site" proposal is inherently a strong-coupling one---it was motivated by the observation that   Ginsparg-Wilson fermions avoid the left-right mixing  that led to a vectorlike spectrum in the strong Yukawa limit of the waveguide model \cite{Golterman:1994at}. The mechanism of the  "one-site" proposal 
 depends on the existence of the strong-coupling symmetric phase of the Higgs-Yukawa theory and on the validity of the "FNN mechanism."

For finite values of $N$ and $a$ we expect that these theories are not equivalent---the global symmetries are different since the 1-site model respects the 133 symmetry, while the warped domain wall model does not. 
In addition, the GW fermion model has a massless gauge mode, while the domain wall model has just a light gauge mode. Thus,  if they are the same it could only be in the intermediate energy regime. 

Which of the two proposed lattice theories is more amenable  to study in practical simulations is a question that we have not touched upon.
We note that, once the two noncompact directions are latticized via Wilson fermions,  the action of our proposed ``warped domain wall" will be manifestly reflection positive. On the other hand, the question of reflection positivity (Hermiticity in real time) of the single-site model deserves further study; to this end, a Hamiltonian formulation might be desirable. Here, we only point out that non-positivity may be irrelevant in the continuum limit---examples of 
non-reflection positive lattice actions appear commonly in lattice constructions of supersymmetric target theories (for  a recent review, see \cite{Giedt:2006pd}). These lattice actions  preserve some exact nilpotent supersymmetries; in fact, demanding invariance under these is the ultimate reason for nonpositivity. Despite non-positivity, however, it has been argued or shown \cite{Giedt:2004qs, Giedt:2004vb} that  in the continuum limit the  models possess  a positive self-adjoint transfer matrix. 

Another issue left for future work is the positivity of the fermion determinant. We have nothing to say about it here and only note that  both the overlap formulation   of ref.~\cite{Narayanan:1994gw} and the construction via Ginsparg-Wilson fermions  of ref.~\cite{Luscher:1998pq} have a complex measure problem:  the Euclidean effective action for 4 dimensional fermions in non-real representations is generally expected to be complex \cite{DellaPietra:1986qd}. 

Finally, we note that  our proposal suggests a way to define the fermion measure for the construction of  \cite{Luscher:1998pq} by obtaining the unbroken chiral gauge theory---our light partition function $Z_{light}$ of (\ref{Zmirror}))---from a particular limit of a vectorlike theory. The fermion measure in our vectorlike models is well defined  and hard questions  of how its phase or   dimensionality change  as one varies the gauge background do not arise. To this end, it would be desirable to understand in more detail the behavior of our ``single-site" model in topologically nontrivial backgrounds.

\appendix
\section{Fermions in continuum and discretized slice of AdS$_3$  } \label{sec:FermionsAdS3}
 
For completeness, in this appendix we present the formulae relevant for the description of fermions in a slice of $AdS_3$ in the continuum and on the lattice (deconstructed version). These expressions are relevant for obtaining the mass matrix (\ref{chargedmassmatrix}) of the charged fermions.

The bulk action for a 3-dimensional Dirac fermion in a slice of $AdS_3$, in terms of  their 2-dimensional Weyl components, is: 
\begin{eqnarray}
\label{bulkcontinuumaction}
S_\Psi^{bulk} &=& \int d^2 x \int\limits_R^{R^\prime} dz \left( \frac{R}{z}\right)^2 \left[ 2 i \bar\psi_- \partial_+\psi_- + 2 i \bar\psi_+ \partial_- \psi_+ \right. \nonumber \\
&& \left. - \frac{i}{2} \left(\bar\psi_+ \partial_z \psi_- - \partial_z \bar\psi_- \psi_+ + \bar\psi_- \partial_z \psi_+ - \partial_z \bar\psi_+ \psi_-\right) \right. \nonumber \\
&& \left. + i \frac{m R}{z} \left(\bar\psi_- \psi_+ - \bar\psi_+ \psi_-\right)\right]~,
\end{eqnarray}
where $m$ is a real mass parameter (odd under 3d parity) and $R$ is the $AdS_3$ curvature radius (the contribution of the spin connection is ``hidden" and can be recovered upon integrating by parts some of the $z$-derivatives).
The equations of motion
are trivially solved for the  fermion zero modes
($\partial_+ = \partial_- = 0$)  and yield two  solutions of opposite chirality:
\begin{eqnarray}
\label{zeromodes}
\psi_-^{(0)} &=& c_1 z^{1 - m R}~, \nonumber \\
\psi_+^{(0)} &=& c_2 z^{1+ m R}~.
\end{eqnarray}
Clearly, the zero modes  (\ref{zeromodes}) can be localized anywhere with the right choice of $mR$. To see that, let us substitute the $\psi_-$ zero mode (and set $\psi_+ =0$)  into  the action (\ref{bulkcontinuumaction}):
\begin{eqnarray}
\label{psi1zeromodeaction}
\psi_-^{(0)} = \chi_-(x) \left(\frac{z}{R}\right)^{1 - mR}~,
\end{eqnarray}
where $\chi_-(x)$ is now an $x^{0}, x^1$ dependent function---the wave function of the zero mode; note that $\chi(x)$ has the same mass dimension (one) as $\psi$. Clearly, only the $\bar\psi_- \partial_+ \psi_-$ term in (\ref{bulkcontinuumaction}) contributes, giving the following 2d action for $\chi_-(x)$: 
\begin{eqnarray}
\label{psi1zeromodeaction1}
S_{c_1} &=& R  \int d^2 x \; 2 i\; \bar\chi_-(x) \partial_+ \chi_-(x) \int\limits_1^{R^\prime/R} d y y^{-2mR}  \\
&=&\frac{R\left(R^{\prime \; 1-2 mR} - R^{1-2mR}\right)}{1- 2 mR} \int d^2 x \; 2 i \; \bar\chi_-(x) \partial_+ \chi_-(x)~.\nonumber
\end{eqnarray}
We  interpret (\ref{psi1zeromodeaction1})  by taking various limits: if $m R > 1/2$, we can certainly take $R^\prime \rightarrow \infty$, i.e. the IR brane to infinity,  and still have a finite action 2d mode. This implies that the $\chi_-$ zero mode is localized near the UV brane if $mR>1/2$. If $mR<1/2$ the localization is nearer the IR brane (this is clear from (\ref{zeromodes}): for, say, negative $mR$, we have $\psi_-^{(0)}$ growing at large-$z$, indicating IR localization). 

Clearly, the story is the opposite for the $\psi_+^{(0)}$ zeromode---the two cases simply differ by the sign of $mR$---so when $mR< -1/2$ (regime where $\psi_-^{(0)}$ was IR-localized) we have UV localization of $\psi_+^{(0)}$. Conversely,   when $mR>-1/2$, $\psi_+^{(0)}$ is IR-localized.
 
Next, we  discretize the following continuum action, obtained from (\ref{bulkcontinuumaction}) upon integration by parts and dropping of the boundary terms---this is needed, as in \cite{Bhattacharya:2005xa}, in order to obtain Wilson terms and hence no doublers in the discretized bulk: 
\begin{eqnarray}
\label{fermionSbulk2}
S_\Psi^{bulk} &=& \int d^2 x \int\limits_R^{R^\prime} dz \left( \frac{R}{z}\right)^2 \left[ 2 i \bar\psi_- \partial_+\psi_- + 2 i \bar\psi_+ \partial_- \psi_+ \right. \nonumber \\
&-& \left.  i \left(\bar\psi_+ \partial_z \psi_- - \partial_z \bar\psi_- \psi_+ \right)  \right. \\
&+& \left. i \; \frac{m R - 1}{z} \left(\bar\psi_- \psi_+ - \bar\psi_+ \psi_-\right)\right]~. \nonumber
\end{eqnarray}
Now we have $N$ 2d Dirac fermions $(\psi_-^k, \psi_+^k)^T$, $k=1,\ldots N$, each charged under the corresponding gauge group, $\psi^k_\pm \rightarrow g_k \psi^k_\pm$; $\partial_\pm \rightarrow D_\pm = \partial_\pm + i A_\pm^{k}$.
Gauge invariant ``hopping" terms between the different groups can be written using the unitary bifundamental links $U_k$, $k = 1, \ldots N-1$.

 The bulk lattice fermion lagrangian we thus  (note that $\partial_z$ in (\ref{fermionSbulk2}) is replaced by the symmetric lattice derivative) obtain is:
\begin{eqnarray}\label{latticefermion2}
L_{\Psi}^{bulk}& = & i\; \sum\limits_{k=1}^N a R w^{k} \left[ 2 \bar\psi_-^{k} D_+\psi_-^{k} + 
2  \bar\psi_+^{k} D_- \psi_+^{k} \right]  \nonumber  \\
& -&  \sum\limits_{k=1}^{N-1}   w^{2 k+1} \left( \bar\psi_+^{k+1} U_k \psi_-^{k} - \bar\psi_-^{k } U_k^\dagger \psi_+^{k+1}\right)        \\
&-& \sum\limits_{k=1}^{N}  w^{2k-1} (1 -  a w m R) \left( \bar\psi_-^{k} \psi_+^{k} - \bar\psi_+^{k} \psi_-^{k} \right) \nonumber .
 \end{eqnarray}
Similar to the gauge field case, we define new  fermion fields:
\begin{eqnarray}
\label{newfermions}
\psi_\pm^k \rightarrow \psi_\pm^k \;  \left( 2 w^{k} a R\right)^{-\frac{1}{2}}~,
 \end{eqnarray}
and the complex conjugate for the $\bar\psi$ fields.
The new fields  now have proper canonical dimension $1/2$. We also define:
\begin{eqnarray}
\label{alpha}
\alpha \equiv 1- a w m R~,
 \end{eqnarray}
in terms of which the new bulk lagrangian is:
\begin{eqnarray}
\label{latticefermion3}
L_{\Psi}^{bulk}& = & i\; \sum\limits_{k=1}^N \;    \bar\psi_-^{k} D_+\psi_-^{k} + 
 \bar\psi_+^{k} D_- \psi_+^{k}     \nonumber   \\
&- &   \sum\limits_{k=1}^{N-1} \frac{w^{k+{\frac{1}{2}}}}{  2a w R} \left( \bar\psi_+^{k +1} U_k \psi_-^{k} - \bar\psi_-^{k} U_k^\dagger \psi_+^{k+1}  \right) \nonumber  \\
&-&    \sum\limits_{k=1}^N w^{k-1}\; \frac{ \alpha}{2awR} \left( \bar\psi_-^{k} \psi_+^{k } - \bar\psi_+^{k } \psi_-^{k } \right)   ~.
 \end{eqnarray}
The mass matrix of eqn.~(\ref{chargedmassmatrix}) can be then easily read off eqn.~(\ref{latticefermion3}). It is also possible to use 
$L_{\Psi}^{bulk}$ to find analytically the zero modes in the discretized version and show that they are well approximated by the continuum expressions (\ref{zeromodes}) at large $N$.

\section{Domain Wall $\beta$-function} \label{sec:BetaFunc}

Our warped domain wall construction arising from $AdS_3$ has the spectrum of fermions that we expect for a chiral gauge theory.  As explained in section~\ref{sec:Fermions} the tower of KK modes becomes heavy and the light neutral modes decouple in the $N \to \infty$ limit.  Since this theory is perturbative in the large $N$ limit, we expect na\"ive decoupling arguments to hold for individual modes.

One might wonder, though, if the large number of modes could have a non-trivial contribution even in the IR.  However, the masses of most of the KK modes for our deconstructed $AdS_3$ are given approximately by:
\begin{equation}
m_n = {\cal O}\left(\frac{w^{-n}}{a R'} \right)
\end{equation}
Only the lightest and heaviest handful of modes deviate from this expression (this spectrum of modes is different from the 3-dimensional continuum where the spacing is linear in $n$).  These masses are rising exponentially, and so we expect that the contribution to an IR propagator from all $N$ modes is not much larger than the contribution from just one mode at the KK scale: $1/R'$.

To verify this expectation, we outline here a 2-dimensional continuum calculation of the $\beta$-function for our theory in the IR using all of the fermions in the entire mass matrix.  We may write our fermion Lagrangian as
\begin{eqnarray}
\frac{i}{2}
\left(\vec{\Psi}^\dagger_-,\vec{\Psi}^\dagger_+\right)
\left(\begin{array}{cc}
D_+ & -\tilde{M}^T \\
\tilde{M} & D_-
\end{array}\right)
\left(\begin{array}{c}
\vec{\Psi}_- \\
\vec{\Psi}_+
\end{array}\right),
\end{eqnarray}
where $\tilde{M}$ is related to the full fermion mass matrix given in equation~(\ref{PsiMass}).  The only difference being that some rows were interchanged since our action here is written with $\Psi^{\dagger}$ rather than $\Psi^T$. The appropriate covariant derivative is $D_\pm=\partial_\pm - i \hat{g}\hat{f}A_{\pm,0}(x)$, where $A_{\pm,0}(x)$ is the 2 dimensional wave function of the lightest gauge mode, $\hat{g}$ is the charge matrix, diagonal with either the site charges $g_i$ for charged modes or zero for neutral modes, and the matrix $\hat{f}$ has the wavefunction of the gauge boson zero mode down it's diagonal entries: $\hat{f}_{i,j} = f_i \delta_{i,j}$, so that these factors together reproduce the gauge coupling of the lightest mode. We are only interested in the lightest gauge mode since we want the one-loop beta function at low energies.  The corresponding wavefunction is given in equation~(\ref{eqn:zeroModeWavefunction}), or it can be found from the gauge boson mass matrix.

The momentum space fermion propagator can be written as:
\begin{eqnarray}
G(p) = -2i
\left(\begin{array}{cc}
p_+ & -\tilde{M}^T \\
\tilde{M} & p_-
\end{array}\right)^{-1}.
\end{eqnarray}
The one loop correction to the $A_+A_+$ correlator is:
\begin{eqnarray}
\int d^2p \textrm{Tr} \left[
\left(\begin{array}{cc}
\frac{\hat{g}\hat{f}}{2} & 0 \\
0&0
\end{array}\right)
G(p)
\left(\begin{array}{cc}
\frac{\hat{g}\hat{f}}{2} & 0 \\
0 & 0
\end{array}\right)G(q-p)\right],
\label{eqn:betaFuncLoop}
\end{eqnarray}
and similarly for the $A_-A_-$ or $A_+A_-$ correlators.  These are all well defined matrices and may be manipulated numerically.  More specifically, we can  plot the momentum dependence of the integrand for the one loop correction in the IR and verify that it is what we expect for a chiral theory.  Only the light left handed modes: $3_-$ and $4_-$, should contribute to the $A_+A_+$ correlator and only the light right handed mode: $5_+$ should contribute to $A_-A_-$.  The $A_+A_-$ correlator only gets contributions from the massive modes  and so it should be suppressed by the KK scale (whatever regulator is used to define our formal continuum 2 dimensional perturbative expansion also contributes to the $A_+A_-$, with a coefficient that can be determined solely by demanding gauge invariance in the anomaly-free theory, hence we need not specify it; see, e.g., the calculation of the 2 dimensional anomaly in \cite{Alvarez-Gaume:1983ig}). 

Since the ``345'' theory is anomaly free, the contribution from the left and right handed modes to  their respective $A_+A_+$ and $A_- A_-$ correlators will be equal;  
it is  therefore sufficient to  consider the anomalous theory of subsection~\ref{sec:Fermions}, where there was only one light charged field, $l_-$.  By numerically calculating and plotting the momentum dependence of the integrand of equation~(\ref{eqn:betaFuncLoop}) we see that there is a pole at both $p_+ = 0$ and $q_+ - p_+ = 0$.  The coefficient of this pole is in fact approximately $g_2^2$, the charge of the light fermions under the restored gauge symmetry.
The related expressions for the $A_-A_-$ and $A_-A_+$ correlators do not show any momentum poles.  We therefore conclude that this theory gives the appropriate chiral $\beta$-function in the IR.

 \bigskip

\acknowledgments
We would like to thank Joel Giedt for contributions to the early part of this work and Yuri Shirman for insightful discussions.
T.B. and M.M. are supported in part by the U.S. Department of Energy under contract W-7405-ENG-36. E.P. acknowledges support by the National Science and Engineering Research Council of Canada (NSERC).

\end{document}